# Gait Change Detection Using Parameters Generated from Microsoft Kinect Coordinates

**Behnam Malmir, Shing I Chang**

*Department of Industrial and Manufacturing Systems Engineering, Kansas State University*
*Manhattan, KS 66506, USA*

## Abstract

This paper describes a method to convert Microsoft Kinect coordinates into gait parameters in order to detect a person's gait change. The proposed method can help quantify progress of physical therapy. Microsoft Kinect, a popular platform for video games, was used to generate 25 joints to form a human skeleton, and then the proposed method converted the coordinates of selected Kinect joints into gait parameters such as spine tilt, hip tilt, and shoulder tilt, which were tracked over time. Sample entropy measure was then applied to quantify the variability of each gait parameter. Male and female subjects walked a three-meter path multiple times in initial experiments, and their walking patterns were recorded via the proposed Kinect device through frontal plane. Time series of the gait parameters were generated for subjects with and without knee braces. Sample entropy was used to transform these time series into numerical values for comparison of these two conditions.

## Keywords
Gait Parameters, Change Detection, Physical Therapy, Microsoft Kinect, Sample Entropy.

## 1. Problem Statement
Gait features analysis, used to study and analyze people's walking patterns, provides essential information for measuring progress of physical therapy. Microsoft Kinect has been proposed to record joint coordinates of the human skeleton in three-dimensional (3D) space; researchers have used these 3D kinematic measures in clinical gait analysis studies. Tupa *et al.* [1] used Kinect to estimate leg length, normalized average stride length, and gait velocity of an individual. These gait features were then compared in three sets of individuals to recognize Parkinson's disease. Vernon *et al.* [2] examined the test-retest reliability measures of some other kinematic measures, such as step length and stride length, to determine whether they can improve prediction performance in common clinical tests.

Quantification is essential for tracking joint health, especially for individuals who are undergoing physical therapy and are affected by an age-related disability. This quantification strategy can be extended to other physical health-related areas. For example, it can help monitor elevated risk of falling as reflected in gait changes due to physical weakness.

This study aimed to detect changes in walking patterns over time using human skeleton coordinates that were converted into gait parameters such as spine tilt, hip tilt, and shoulder tilt. The proposed procedure can be applied to fall prediction of elderly people, physical therapy, and sport science. Similar studies have shown the potential of Microsoft Kinect for fall risk [3] assessment, as well as clinical and field-based assessment of gait [4-6], such as Timed Up and Go (TUG). A review of technical and clinical impacts of Kinect on physical therapy and rehabilitation is included in [7], and other issues, such as sleep disorders [8] or recognition of breathing [9], were analyzed using Kinect.

## 2. Kinect Operation
Microsoft Kinect was initially designed to enhance the video gaming experience by capturing a gamer's joint position. Consequently, Kinect presents a simple, inexpensive, and portable method of examining motion in a human subject test, such as TUG, without intrusion on human subjects [2].

Kinect is unique and useful for gait analysis because it contains an RGB camera, a depth sensor, and a multiarray microphone. The Kinect's depth sensor can capture 3D data and does not require of lighting systems, allowing it to capture data indoors or outdoors. Kinect possesses a frame rate of 9–30 fps and a resolution of 640 x 480 that can be increased to 1280 x 1024 using a lower frame rate [7]. In this study, Kinect was set to record 30 fps. Effectiveness of Kinect for gait analysis was tested in the study by Tupa *et al.* [1], achieving an accuracy of 97.2% and suggesting potential use of Kinect image and depth sensors for the aforementioned applications. However, experiment results from this study showed difficulty in direct detection of gait changes using raw data captured by Kinect. Figure 1 depicts eight selected joints of a female subject walking three meters. The proposed solution overcomes this problem by converting coordinate data into gait parameters.

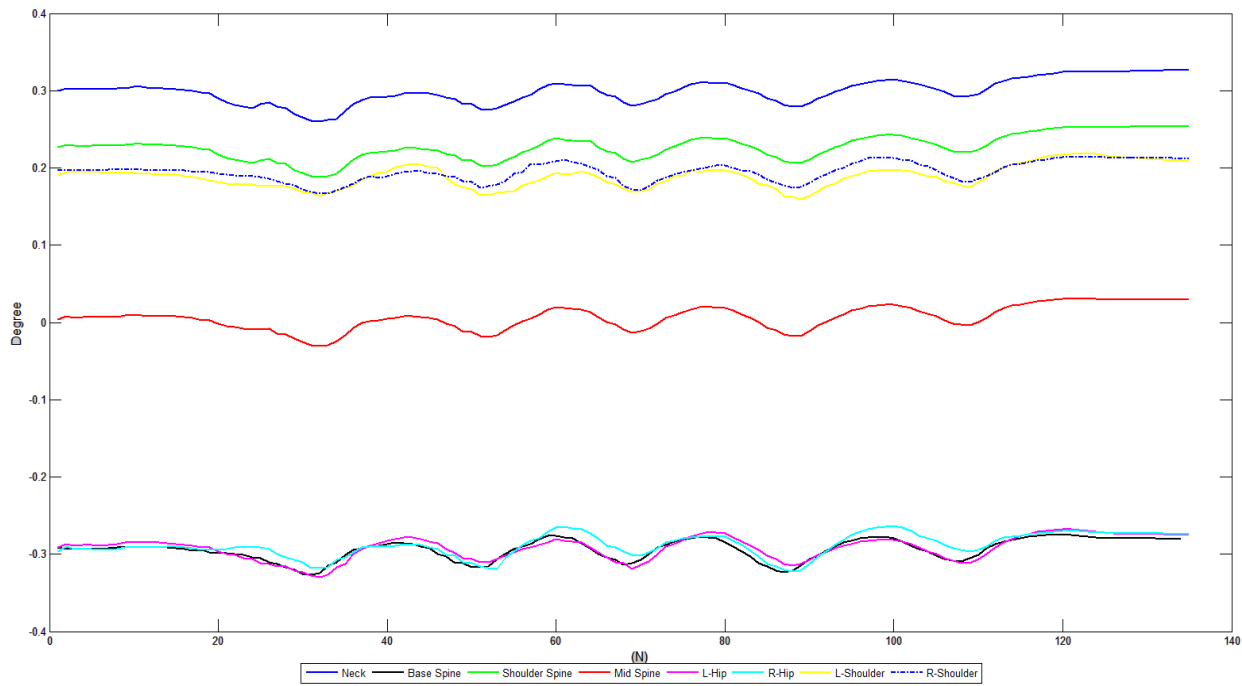

Figure 1: Tracking eight selected Kinect joints of a healthy female subject in Y coordinate over time

## 2.1 Skeleton Modeling

In this study, Microsoft Kinect was used to track 25 joints over time. Although those joints formed a human skeleton, only some of the joints were relevant to shape gait parameters that were proposed to detect gait changes. Table 1 lists relevant joints generated by Kinect with corresponding gait parameters through the frontal plane [10].

Table 1: Descriptions of gait parameters

| Gait Parameter | Assigned as | Relevant Joints |
|---|---|---|
| Spine Tilt | $V_1$ | Shoulder Spine (ShS) - Base Spine (BS) |
| Hip Tilt | $V_2$ | Left and Right Hips (LH-RH) |
| Shoulder Tilt | $V_3$ | Left and Right Shoulders (LSh-RSh) |

X and Y coordinates generated from Kinect joints ShS, BS, LH, RH, LSh, and RSh listed in Table 1 can be converted into gait parameters such as spine tilt, hip tilt and shoulder tilt. The X dimension typically tracks the axis perpendicular to the camera's line of sight, whereas the Y dimension tracks the up-and-down movement of a person; Y and Z coordinates of the joints may also be considered in cases where Kinect is used from the Sagittal perspective. Z direction tracks if a subject is closing in or fleeing the camera. In this study, however, only X and Y directions were necessary.

## 2.2 Gait Parameter Processing

In anatomy, hip tilt is the orientation of the pelvis in respect to the femurs it rests upon and in space. Hip tilt, spine tilt, and shoulder tilt of 5 degrees each is typical for walking, meaning that the respective tilts should be no more than 5 degrees for a healthy subject. Deviation over 5 degrees indicates that a subject may suffer a physical problem in walking properly.

All subjects in this study were healthy. In fact, one of the study goals was to analyze the measured tilt of their spines, hips, and shoulders during walking and compare those measurements to other healthy subjects. In addition, differences in tilt measurements on the same subject with and without a knee brace on his right knee were analyzed. Gait parameters and relevant joints are illustrated in Figure 2.

The mathematical relation between relevant joints in order to acquire gait parameters in the frontal plane is

$$y_2 - y_1 = m(x_2 - x_1), \qquad (1)$$

where $(x_i, y_j)$, $I = 1,2$ is the coordination of two selected Kinect joints that were considered to be the connection vectors and were tracked over time. Specifically, $x_1$ and $x_2$ were assigned as the X coordinates of the left- and right-sided joints, and $y_1$ and $y_2$ were assigned as the Y coordinates of the left- and right-sided joints, respectively. Parameter $m$ is the slope of these vectors, changing over time. Then $m$ was converted to the angle between two joints at each point by the relation $m = \tan\alpha \Leftrightarrow \alpha = \text{Arctan}\, m$, where $\alpha$ is a radian-based measure. In order to convert $\alpha$ to degree, we multiplied it by $180/\pi$. MATLAB (R2014a) was used for computation.

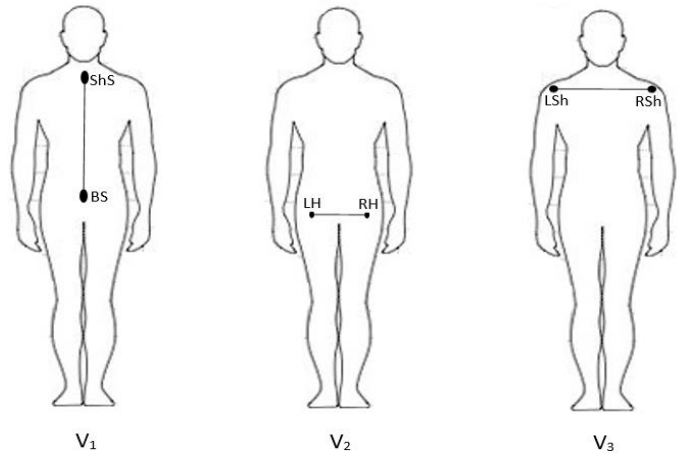

Figure 2: Relevant joints of proposed gait parameters

## 3. Experimental Design and the Proposed Method

### 3.1 Statistical Procedures

One of the goals of this study was to differentiate two subjects' movements based on their Kinect coordinates. Data were derived from Microsoft Kinect coordinates by recording 25 joints of the human body over time. Once Kinect coordinates were converted into gait parameters considering their relevant joints, the tilts (i.e., slope between two relevant joints) were measured over time, and then angle or tilt changes over time were examined using Sample Entropy (SE) measure [11, 12]. In this study, SE was applied to quantify trends and detect pattern changes in a person's gait.

In order to demonstrate operation of the proposed method, this study was conducted in two main steps. First, walking patterns were recorded from two subjects via the proposed Kinect device, as described in Section 3.2, and the data was used to determine whether or not their walking patterns differed. As described in Section 4, possible differences in walking patterns were examined using the same set of gait parameters when a subject wore a knee brace on his right knee.

Although regression models may be useful for finding significant relationships between gait parameters and factors such as age, gender, height, and weight, this study demonstrated that two subjects of the same gender, age, and approximate weight and height showed differences in their walking patterns. The experimental results showed promise of using gait parameters over time to track progress or lack of in successive physical therapy sessions.

### 3.2 Experiments and Data Collection

In the first experiment of this study, two similar subjects were compared. The subjects chosen were completely healthy with similar physical conditions such as age, gender, height, and weight. Both subjects were instructed to wear comfortable shoes and walk in a straight three-meter path. A Microsoft Kinect camera was placed in the frontal direction. Each experimental setting for a gait parameter was repeated 10 times.

Experimental results are shown in Figures 3 and 4, where $v_1$, $v_2$, and $v_3$ represent spine tilt, hip tilt, and shoulder tilt, respectively. Figure 3 compares two young female subjects in three-meter walking experiments (forward and back). Three gait parameters are represented via box plots of 10 replicates. SE was used to summarize the variability of one walking path either forward or back. The notation $W_{ijv}$ indicates the distribution of 10 replicates for each experimental setting according to

$$W_{ijv} \begin{cases} \text{Subjects} & i \in (1,2) \\ \text{Forward and Back} & j \in (1,2) \\ \text{Parameters} & v \in (1,2,3) \end{cases}$$

Sign of $\otimes$ indicates average of SE values based on 10 replicates; the horizontal bar represents the median.

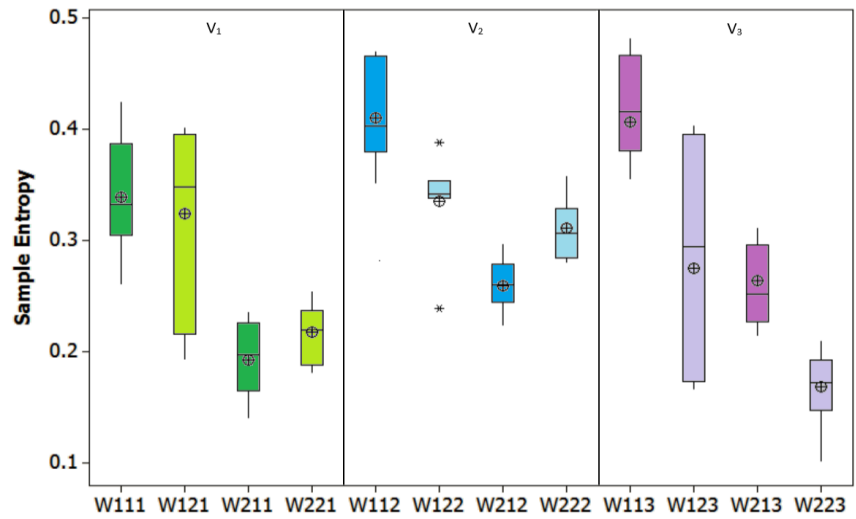

Figure 3: Measure of variability using SE of three gait parameters on two subjects

In order to compare variabilities between the two subjects, $W_{1jv}$ must be compared to $W_{2jv}$. For example, SE values of Subject 1 differed from SE values of Subject 2 in terms of spine tilt parameter ($V_1$) when the subjects walked forward, as shown in the first boxplot ($W_{111}$) and the third box plot ($W_{211}$) in Figure 3. Similarly, other pairs could also be compared. Results showed that Subject 1 walked differently than Subject 2.

SE application to one time series that was demonstrated in [12] could be expanded into multiple channels that track Kinect joints, as shown in Figure 1. Figure 4 compares movement profiles of the two subjects based on three gait parameters using star glyphs. In the figure, the first subject is labeled $S_1$ and the second subject is labeled $S_2$. This figure provides an efficient visual comparison to contrast differences between the two subjects using three gait parameters. Many differences between the two subjects during walking forward or walking back are evident.

## 4. Gait Change Detection

This section investigates whether the proposed method can detect a person's gait change, considering all proposed gait parameters. We propose to use the gait parameters to identify whether there is any change in a subject's walking patterns with and without wearing a knee brace. The same three-meter walking test was conducted multiple times on a healthy 25-year-old male subject to detect possible changes in his walking pattern using Kinect. The subject walked forward normally, through the three-meter path one time wearing a brace on his knee (first test) and another time without the brace (second test). This procedure was replicated 10 times.

Skeleton information of the subject was generated by the Kinect skeleton feature, and then the SE measure on each gait parameter was obtained. Each test replication contained three gait parameters ($V_1$, $V_2$, and $V_3$) for characterizing the physical status of the subject. Final results are shown in Figure 5, where the first test is defined as $T_1$ and the second test is defined as $T_2$. These figures provide insight into the possible changes.

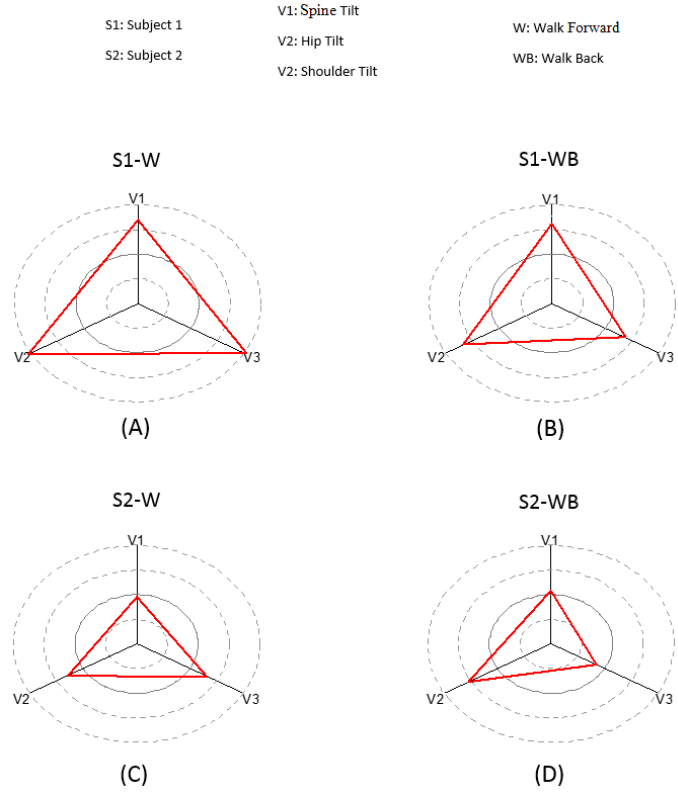

Figure 4: Comparison of the mean SE values of the profiles using three gait parameters of Subject 1 (A, B) and Subject 2 (C, D) during walk-forward and walk-back experiments.

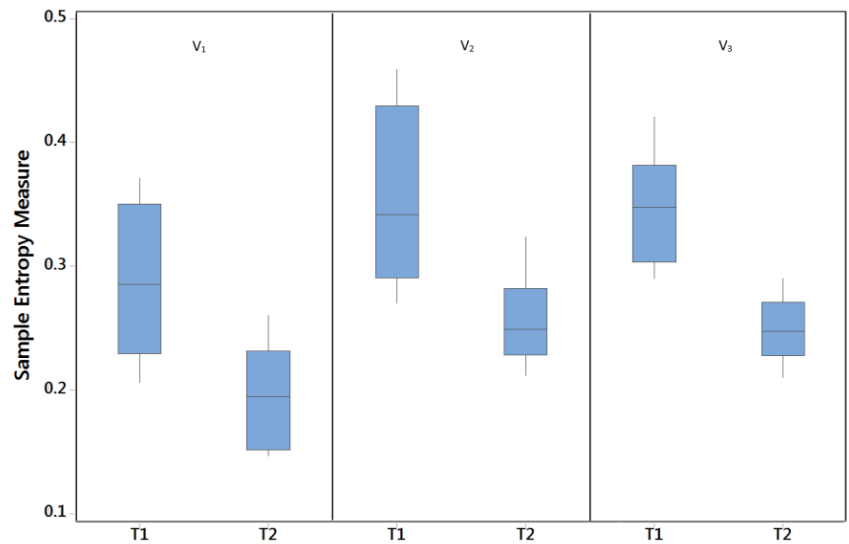

Figure 5: Comparison of a subject's walking status in two tests (T1: with a knee brace; T2: without a knee brace) using SE values of $v_1$: spine tilt, $v_2$: hip tilt, and $v_3$: shoulder tilt over time.

Figure 5 shows prominent differences between SE values of each gait parameter for one person in two conditions. The variability of SE values in the first test (T1) was larger than the variability of SE values in the second test (T2). Therefore, any gait parameter introduced in Section 2.1 may be a potential candidate to measure physical therapy progress. However, combinational use of all three gait parameters on star glyphs may provide a more concise presentation of gait changes if tracked over time.

## 5. Conclusion and Ongoing Studies

This study demonstrated the potential use of human kinematic measurements in clinical gait analysis. Coordinate measures were initially converted into gait parameters related to walking patterns. Then, three gait parameters were analyzed to compare walking patterns of individuals with identical physical conditions. Initial experimental results demonstrated that the proposed method is capable of detecting walking pattern differences in individuals. The proposed SE (Sample Entropy) measure was used to summarize data from each walking path into one value.

The proposed gait parameters are capable of detecting gait changes in one person based on the data collected from the frontal plane. SE distributions collected from the test subject showed that all three gait parameters have different values in two different conditions (walking with and without wearing a knee brace). This results show potential for use to track the progress (or the lack of progress) in successive physical therapy sessions. Revised Sample Entropy [13] could be applied to increase procedure sensitivity for detecting changes. However, additional test subjects are needed to improve the reliability of this study. Other statistical techniques may also be used in lieu of the SE since SE may not be familiar to most statisticians. Finally, the proposed procedure in this paper could be coupled with statistical process control techniques to monitor possible elevated fall risks of elderly people.